# KIC 8462852: Potential repeat of the *Kepler* day 1540 dip in August 2017

Rafik Bourne[1] and Bruce Gary[2]

[1]East Cannington, Perth, Western Australia 6107

[2]Hereford Arizona Observatory, Hereford, AZ 85615

**Abstract**

We report 33 V-band observations by the Hereford Arizona Observatory (HAO) of the enigmatic star KIC 8462852 during the two week period 3-17 August 2017. We find a striking resemblance of these observations to the *Kepler* day 1540 dip with HAO observations tracking the *Kepler* light curve (adjusted for egress symmetry). A possible explanation of this potential repeat transit is a brown dwarf and extensive ring system in a 1601-day eccentric orbit. We suggest this object may be detectable through radial velocity observations in October and November 2017, with an amplitude of ~ 1-2 kms$^{-1}$.

E-mail: blgary@umich.edu

The enigmatic star KIC 8462852 (hereafter KIC8462) was found to vary in brightness during the 4 years of the *Kepler* mission by up to 22% with no obvious periodicity (Boyajian *et al.*, 2016).

This Research Note reports on a potential repeat interval of 1601-days of the *Kepler* day 1540 (D1540) dimming event in observations by the Hereford Arizona Observatory (HAO) during August 2017.

Gary and Bourne, 2017 (hereafter GB17), suggested a 1600-day repeat of a yearlong U-shaped fade in KIC8462, seen in *Kepler* full frame images (reported by Montet and Simon, 2016) and also present in HAO data in 2017. In addition, GB17 point out that most dips occur during the later part of the U-shaped fades. Support for the U-shaped fades comes from Simon et al. (2017), who report a temporary increase in the brightness at KIC8462 after the *Kepler* mission, which is based on ASAS data. A repeat of a yearlong U-shaped fade in 2008 is not excluded by ASAS data.

Sacco et al. (2017) interpreted the 2017 dips as a repeat of the 2013 dips with a period of 1574 days by correlating dips in the 2013 and 2017 light curves.

Figure 1 presents the normalized flux for D1540 as well as a reversed display of egress overlaid on to the ingress portion of the graph. The reversed egress has been "stretched" by a factor of 1.08 to align the egress and ingress periods. This small adjustment is made to correct for the asymmetry between ingress and egress (due to a likely increasing transit velocity) during the 18 day transit caused by orbit eccentricity.

Initial transit simulation modeling for D1540, undertaken in 2016, suggested the transiting object may be a brown dwarf (BD) with a ring system in an eccentric orbit. A video of this simulation is at https://youtu.be/4fQfaMxc9lE. The simulation modeling of D1540 is the subject of a paper in preparation by Bourne, Gary and Plakhov, and is suggestive of a transit velocity ~ 25-30 kms$^{-1}$. If this broad transit feature is a massive ring system, the system would be ~ 0.2 AU in diameter extending well beyond the Roche limit of a BD and would therefore be transient.

The HAO observation of KIC8462 during the two week period 3-17 August 2017 recorded a reduction in flux of up to 1.5% in V-band (Normalised Flux 1 = mag 11.9077). A comparison of the 33 hourly binned HAO observations with D1540 (shifted 1601 days) shows 25 of the observation aligns with D1540 egress (and the reverse egress curve for ingress period) within the HAO hourly bin error limits of ~ 0.1% (1 mmag).

The observation on JD 2457974 was made by Dr Boyajian using the Las Cumbres Teide Observatory in Tenerife, Canary Islands, Spain (LCO - TFN), in the r'-band (from Dip Update 53n, WTF blog http://www.wherestheflux.com/single-post/2017/08/09/Dip-update-53n).

We note:
1) 25 of 33 HAO observations in August 2017 match the *Kepler* data (and reverse egress) from 4.4 years earlier with a precision of ~ 0.1%, and
2) the LCO TFN observation matches the brief central dip present in D1540, shifted by 1601 days.

The next transit of this object, in a 1601-day orbit scenario, is expected on 27 December 2021.

Ballesteros et al. (2017) found the standard deviation of the *Kepler* light intensities (one-week periods) exhibits a minimum near *Kepler* day 800, when our modeling of the hypothetical BD orbit places the object farthest from the star, and is maximum at a time that the model places the object closest. This indirectly supports the suggestion of a 1601-day period and the orientation for the proposed object's eccentric orbit.

An eccentric 1601-day orbital period for a substantial object, possibly a BD with an extensive ring system, is consistent with:

- the suggested 1600-day period for the one year U-shaped fades,
- the occurrence of dips in the latter part of the U-shaped fades,
- light intensity variations peaking at the beginning and end of the *Kepler* observing period with a minima near day 800,
- a possible repeat of D1540 in August 2017, and
- an apparent increase in transit velocity during D1540/August 2017 being consistent with an eccentric orbit as the object approaches periapsis.

The presence of a BD in an eccentric 1601-day orbit could be detectable in October and November 2017 through radial velocity (RV) observations with a potential amplitude of ~ 1-2 kms$^{-1}$.

ACKNOWLEDGMENTS

We thank Dr T. Boyajian for making LCO observations available in the public domain, which allowed us to learn that the August 2017 dimming event exhibited a brief and deep central dip on JD 2457974 and for helpful comments during the drafting of this RN.

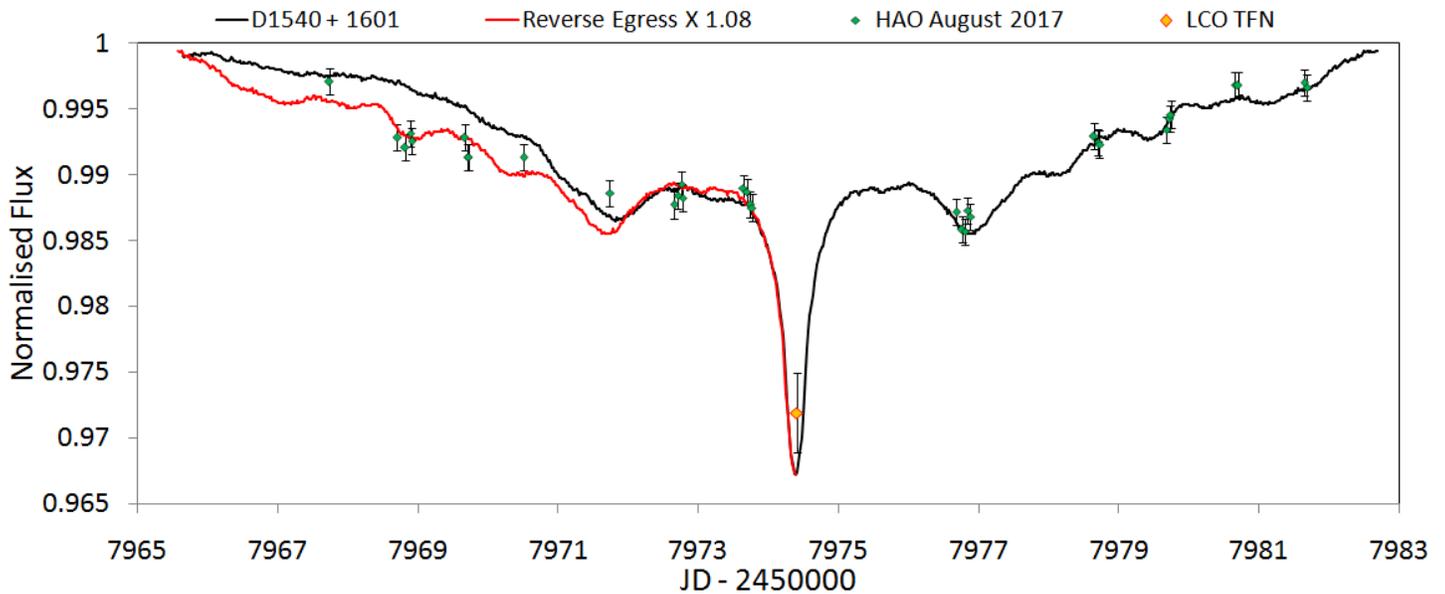

**Figure 1.** Kepler D1540 shifted forward 1601 days (black trace), HAO August 2017 V-band data (green diamonds) and a LCO TFN r'-band observation on JD2457974 obtained by Dr Boyajian (orange diamond). The red trace is a reverse of the *Kepler* egress observations.